\documentclass{article}
\usepackage{spconf,amsmath,graphicx,hyperref, xcolor}
\usepackage{booktabs}
\usepackage{makecell}
\usepackage{multirow}
\usepackage{amssymb, bm}
\usepackage{subfig}
\usepackage{cite}
\usepackage{float}

\newif\ifmk
\mktrue

\title{Domain-Incremental Learning for Generative Speech Enhancement}
\name{Manjunath Mulimani$^{1}$,
       Annamaria Mesaros$^{2}$,
       Minje Kim$^{3}$,      
       Jesper Rindom Jensen$^{1}$
       \thanks{We acknowledge the EuroHPC supercomputer LUMI, hosted by CSC (Finland) and the LUMI consortium.}}
 \address{ $^1$ Aalborg University, Department of Electronic Systems, Denmark\\
 $^2$ Tampere University, Signal Processing Research Centre, Finland \\
  $^3$ University of Illinois Urbana-Champaign, Siebel School of Computing and Data Science, USA 
\\
  }
\begin{document}
\ninept
\maketitle
\begin{abstract}
We propose a domain-incremental learning framework for generative speech enhancement  (SE) that learns from a sequence of datasets or domains recorded under diverse acoustic conditions.  Fine-tuning a pretrained model on continuously evolving domains leads to catastrophic forgetting of previously acquired knowledge, while zero-shot generalization often fails to adequately adapt to unseen domains. To address these challenges, we first develop a novel language model-based generative SE model that we then use as a pretrained backbone and incrementally adapt it to acoustically mismatched domains using lightweight domain-specific Low-Rank Adaptation. The proposed framework enables the model to acquire enhancement capabilities for new domains while preserving performance on previously learned domains. 
Evaluated on four heterogeneous speech datasets, our approach effectively adapts to new domains without forgetting previously learned domains.
\end{abstract}
\section{Introduction}
\label{sec:intro}
The objective of speech enhancement (SE) is to improve the intelligibility and perceptual quality of speech signals corrupted by adverse acoustic conditions. Existing approaches fall into two broad categories: discriminative and generative. Discriminative models learn a direct mapping from noisy speech to clean speech; however, they often overlook the rich semantic information embedded in speech, which can be crucial for achieving high perceptual quality. 
In contrast, generative models leverage semantic representations to generate more natural and perceptually pleasing speech while effectively suppressing background noise. 

A wide range of generative models have been successfully applied to speech enhancement, including variational autoencoders (VAEs) \cite{fang2021variational}, generative adversarial networks (GANs) \cite{fu2019metricgan}, flow-based models \cite{nugraha2020flow}, diffusion probabilistic models \cite{lu2022conditional}, \cite{lemercier2023storm}, and, more recently, discrete token-based language models (LMs) \cite{yang2024genhancer, wang2024selm, yao2025gense, li2025sense}. However, generalization to unseen domains remains a major challenge for SE systems. Studies have shown that generative speech enhancement models generalize more effectively to unseen domains than conventional discriminative methods \cite{fang2021variational}, \cite{lu2022conditional}. Among these approaches, LM-based generative models have demonstrated particularly better generalization capabilities \cite{yao2025gense}, \cite{li2025sense}.

Despite their architectural differences, most generative SE systems are trained under the assumption of a fixed training distribution. In practice, deployed SE systems encounter 
variations in speakers, background noise, languages, dialects, recording devices, and other acoustic conditions. The zero-shot generalization capability of these models enables them to handle unseen domains to some extent, but 
their performance can degrade substantially when significant domain shifts exist between the training and deployment conditions. In contrast, fine-tuning a model on a new domain  
leads to a deterioration in SE performance on those earlier domains, a phenomenon known as \emph{catastrophic forgetting}. A straightforward solution is to store data from all domains and completely retrain the model whenever new data become available. However, such an approach is often impractical due to privacy concerns, storage requirements, and computational constraints in real-world applications. 

In this work, we aim to develop a domain-incremental learning (DIL) framework for a generative SE system that learns to enhance speech from different domains that arrive sequentially over time.  
DIL differs fundamentally from existing domain adaptation (DA) studies in SE \cite{frenkel2023domain}, \cite{raichle2026test}. DA typically considers two domains: source and target, and aims to transfer knowledge from the source domain to improve performance on the target domain. In contrast, DIL involves a sequence of multiple domains that arrive over time and must be learned incrementally, and evaluates the overall performance across all domains observed so far.

DIL has been successfully applied to sound event classification across different datasets \cite{casciotti2026domain}, acoustic scene classification across different cities \cite{mulimani2025domain}, and speech recognition \cite{fu2021incremental}, \cite{javed25_interspeech}. However, these tasks involve predicting a predefined set of class labels or linguistic units. In contrast,  DIL for SE requires preserving the ability to generate high-quality clean speech across previously encountered domains while adapting to new acoustic conditions, making it fundamentally different and more challenging than classification-based DIL.

\begin{figure}[!tbp]
  \centering
  \subfloat[]{\includegraphics[width=\linewidth, height=0.45\linewidth]{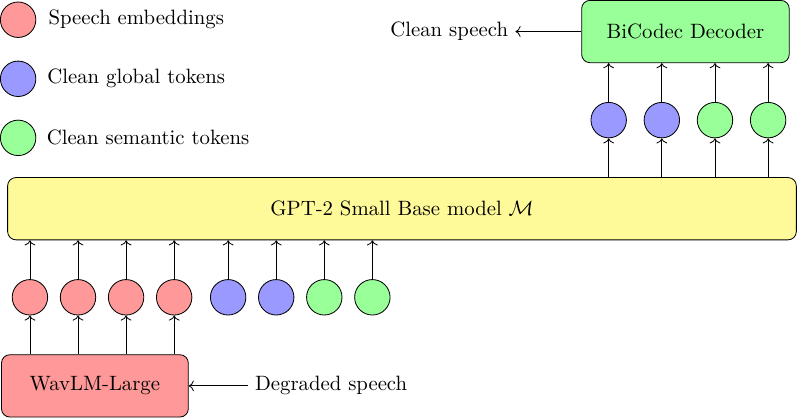}\label{fig:gpt2}}\\
  \vfill  \vspace{-10pt}
  \subfloat[]{\includegraphics[width=\linewidth, height=0.35\linewidth]{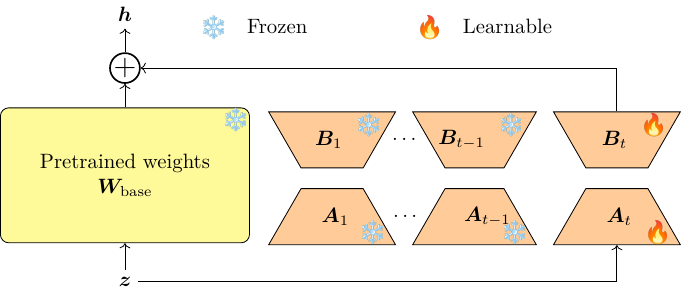}\label{fig:lora}}
  \vspace{-5pt}
  \caption{An overview of the proposed DIL for generative speech enhancement. (a) The base model $\mathcal{M}$ is trained on domain $\mathcal{D}_0$. (b) LoRA is added to $\mathcal{M}$ for each incremental domain $\mathcal{D}_t$. 
  }
     \label{fig:ODFD}
     \vspace{-16pt}
\end{figure}

A few studies have investigated DIL for SE. For example, \cite{lee20d_interspeech} proposes a regularization-based approach to mitigate catastrophic forgetting, whereas \cite{yang2024learning} introduces an adapter-based framework that achieves improved performance in the DIL setting. However, both methods are developed for discriminative SE models. Whether such DIL strategies remain effective for LM-based generative SE models, which formulate enhancement as an autoregressive token-generation task, remains largely unexplored. Furthermore, existing DIL-based SE studies primarily consider domain shifts arising from the introduction of new noise types. In contrast, we investigate DIL for a generative SE model under broader domain shifts arising from different speech enhancement datasets.
To address these limitations, we propose a DIL framework for generative SE.

The main contributions of this work are as follows: 
(1) we develop a GPT-based generative SE model that we train from scratch using discrete semantic and global speech tokens from the Spark-TTS BiCodec tokenizer \cite{wang2025spark} and investigate its generalization and adaptation capabilities across acoustically mismatched domains; 
(2) we propose a domain-incremental learning framework that adapts the pretrained generative SE model to sequentially arriving domains using domain-specific LoRA adapters while preserving domain-shared knowledge in the frozen base model; 
and 
(3) we evaluate the proposed framework on four heterogeneous speech enhancement datasets that exhibit substantial shifts in speech characteristics, speakers, recording conditions, acoustic environments, and noise distributions, providing a benchmark for DIL generative SE.

\section{Method}
\label{sec:method}

\subsection{DIL tasks setup and notations}
In our domain-incremental learning (DIL) setup, we first train a base language model $\mathcal{M}$ for SE on a large-scale dataset $\mathcal{D}_0$. Subsequently, the pretrained model $\mathcal{M}$ is exposed to a sequence of $T$ tasks, defined by additional datasets, $\mathcal{D}_1, \mathcal{D}_2, \ldots, \mathcal{D}_T$, which are available to the model sequentially over time. Each domain consists of its own dataset $\mathcal{D}_t$, which contains noisy speech recordings and their corresponding clean speech signals, collected under acoustic conditions assumed to differ from those of other domains, although a certain level of overlap between the domains is unavoidable.

At each incremental step, $\mathcal{M}$ is trained to enhance speech from a newly arriving domain whose underlying data distribution may mismatch previously observed domains. Importantly, data from previously encountered domains are unavailable during the training of the current domain. During inference, the performance of $\mathcal{M}$ is evaluated on all SE tasks seen so far. Throughout this work, we use the terms \emph{task}, \emph{dataset}, and \emph{domain} interchangeably to refer to $\mathcal{D}_t$.
\subsection{Base model for generative speech enhancement}
The generative SE model autoregressively predicts a sequence of discrete tokens corresponding to the target clean speech signal, conditioned on acoustic embeddings extracted from the degraded speech signal. The generated tokens are subsequently decoded to reconstruct the enhanced speech waveform. Fig.~\ref{fig:gpt2} illustrates the proposed {GPT-2 Small} generative SE base model $\mathcal{M}$, which is trained from scratch to perform this token generation task.

To construct the acoustic embeddings used for conditioning, 
we encode the noisy speech signal using WavLM-Large \cite{chen2022wavlm}, a pretrained speech foundation model.
Specifically, representations extracted from the sixth transformer layer are used as an acoustic prefix, producing a 1,024-dimensional embedding every 20 ms frame.
We tokenize the corresponding clean speech into discrete semantic and global tokens using the Spark-TTS BiCodec tokenizer \cite{wang2025spark}. 
The tokenizer generates 50 semantic tokens per second and a fixed set of 32 global tokens for each speech utterance. 
Semantic tokens are obtained through vector quantization (VQ) using a single codebook of size 8,192. Global tokens are generated by quantizing ECAPA-TDNN \cite{desplanques20_interspeech} speaker embeddings with Finite Scalar Quantization (FSQ) \cite{mentzer2024finite}, which uses six quantization dimensions with four levels per dimension, i.e., $4^6 = 4,096$ as the codebook size.

The model $\mathcal{M}$ is trained to maximize the posterior probability of the clean speech utterance $\bm s$ given its noisy observation $\bm x\in\mathbb{R}^n$: $p(\bm s|\bm x) = \prod_{i=1}^{n} p(s_i \mid s_{1},\ldots,s_{i-1}, g_{1\ldots G}, e_{1\ldots n})$, where $s_i$ denotes the semantic tokens at time index $i$ and $g_k$ denotes the global tokens,  
generated in a similar autoregressive manner:
$p(\bm g|\bm x) = \prod_{k=1}^{G} p(g_k\mid g_{1}\ldots,g_{k-1},e_{1\ldots n})$, where $e_i\leftarrow\mathcal{E}(x_i)$ denotes the embeddings extracted from WavLM-Large $\mathcal{E}$, $n$  is the total number of frames, and $G$ is the total number of global tokens per utterance.  

The vocabulary of $\mathcal{M}$ consists of $4,096+8,192+2$ tokens, corresponding to the global tokens, semantic tokens, and the special \textless Beginning of Sequence\textgreater\ (BoS) and \textless End of Sequence\textgreater\ (EoS) tokens. During training, the input sequence to $\mathcal{M}$ is structured as "\textless BoS\textgreater, embeddings from WavLM-Large (prefix), global tokens, semantic tokens, \textless EoS\textgreater".
Each token is represented using a 768-dimensional embedding in GPT-2 Small. To match this embedding space, the 1,024-dimensional WavLM representations are projected to 768 dimensions before being concatenated with the token embeddings and fed into $\mathcal{M}$. 

During inference, noisy embeddings $e$ together with \textless BoS\textgreater are provided to the model $\mathcal{M}$ as a prefix. 
Conditioned on this prefix, $\mathcal{M}$ first generates a fixed-length sequence of global tokens, followed by frame-level semantic tokens until the \textless EoS\textgreater\;  token is produced. The generated global and semantic tokens are subsequently passed to the BiCodec decoder, which reconstructs the enhanced speech.
 
\subsection{DIL for generative speech enhancement}
We introduce LoRA adapters into the pretrained base model $\mathcal{M}$ for each incremental speech enhancement domain $\mathcal{D}_t$, as illustrated in Fig.~\ref{fig:lora}.
Specifically, LoRA adds low-rank decomposition matrices $\bm A_t\in\mathbb{R}^{d_\text{in}\times r}$ and $\bm B_t\in\mathbb{R}^{r\times d_\text{out}}$ for each domain $\mathcal{D}_t$ to a weight matrix $\bm W_\text{base}\in\mathbb{R}^{d_\text{in}\times d_\text{out}}$ of $\mathcal{M}$ as: $\bm W_\text{base} + \bm{\Delta W_t} = \bm W_\text{base} + \bm A_t\bm B_t$,
where rank $r\ll \min(d_\text{in}, d_\text{out})$.     During training,  $\bm W_\text{base}$ remains frozen and is shared across all domains, while only the lightweight domain-specific LoRA parameters $\bm A_t$ and $\bm B_t$ are optimized for a corresponding adaptation task for the $t$-th domain. For a given input $\bm z$, the original forward pass $\bm h=\bm W_\text{base}\bm z$ is modified as: $\bm h=\bm W_\text{base}\bm z+\bm A_t\bm B_t\bm z$,
where $\bm h$ is the hidden output. This enables the model $\mathcal{M}$ to acquire domain-specific knowledge through a small number of trainable parameters while preserving the domain-invariant knowledge encoded in the frozen base model. 

During inference, the SE performance of $\mathcal{M}$ depends on the choice of the specific LoRA module, while the optimal choice $t^*$ is not known. We propose two selection strategies:: \emph{domain-aware} and \emph{domain-agnostic}. In the domain-aware setting, both the ground-truth domain identity $t^*$ and the test sample are provided to $\mathcal{M}$. The domain identity is used to select the corresponding LoRA parameters, which are then employed to generate the clean global and semantic tokens for the noisy input speech. However, domain information may not always be available in practical deployment scenarios. Therefore, we also consider a domain-agnostic setting, where only the noisy test sample is provided to $\mathcal{M}$. In this case, inspired by \cite{mulimani2025domain}, the most suitable domain-specific LoRA adapter is selected based on the predictive uncertainty of the model. Specifically, the test sample $\bm x$ is forwarded through all LoRA adapters associated with the domains observed so far, producing a probability distribution over the vocabulary for each adapter. We then compute the prediction uncertainty for domain $\mathcal{D}_t$ using the entropy of the predicted distribution:
\begin{equation}
    \mathcal{U}(\mathcal{D}_t) = -\sum_{v=1}^{V} p\left(\bm y^{(\mathcal{D}_t)}_v\mid \bm x\right)\log  p\left(\bm y^{(\mathcal{D}_t)}_v\mid \bm x\right),
    \label{uncertainity}
\end{equation}
where $\bm y$ is the output on $\mathcal{D}_t$ parameters ($\bm W_\text{base} + \bm\Delta\bm W_t$) and  $V$ is size of the vocabulary in $\mathcal{M}$. The LoRA adapter yielding the minimum entropy is selected, i.e., $\hat{t^*}=\text{arg min}_t \mathcal{U}(\mathcal{D}_t)$, as lower entropy indicates higher confidence and lower predictive uncertainty. For brevity, we hereafter refer to the proposed domain-incremental learning framework for generative speech enhancement as \textbf{DIL-GenSE}.

\section{Evaluation and Results}
\label{sec:pagestyle}
\subsection{Datasets, training setup, and baselines}
To construct the domain-incremental learning benchmark we use five datasets recorded under diverse acoustic conditions: 

\noindent  \textbullet \hspace{1pt}
\textbf{$\mathcal{D}_0$ (Base training dataset):} We use English clean speech from the Emilia-Large dataset \cite{he2024emilia}, excluding the Emilia-YODAS split. Approximately 10,400 hours of clean speech with a DNSMOS score greater than 3.40 are selected. To generate noisy-clean speech pairs, we use 180 hours of noise recordings from the DNS Challenge 2020 dataset \cite{reddy20_interspeech}. We synthesize noisy speech at SNR levels from $-5$ to $20$ dB without reverberation. All remaining parameters follow the recommendations of \cite{li2025sense}. 

\noindent \textbullet \hspace{1pt} \textbf{$\mathcal{D}_1$ (DNS):} For incremental adaptation, we use 500 hours of clean speech together with the same 180 hours of noise data used in $\mathcal{D}_0$ to generate 100 hours of noisy-clean speech pairs using the official DNS2020 data-generation script. Although the noise sources are shared with $\mathcal{D}_0$, we use different speech sources to create a domain shift arising from variations in speech characteristics and recording conditions. The training and validation splits follow \cite{kuhne2026mambattention}, while the DNS2020 test set without reverberation is used for evaluation. 

\noindent \textbullet \hspace{1pt} \textbf{$\mathcal{D}_2$ (EARS):} This domain consists of noisy-clean speech pairs from the EARS-WHAM\_v2 dataset. The training, validation, and test splits are selected according to the protocol described in \cite{kuhne2026mambattention}. 

\noindent \textbullet \hspace{1pt} \textbf{$\mathcal{D}_3$ (LibriTTS-R):} We use 100 hours of speech from the LibriTTS-R dataset for training and select 5,000 utterances from the remaining 360-hour portion for validation. The test set contains 4,837 utterances. Noise recordings are drawn from the TUT Urban Acoustic Scenes 2018 Development dataset \cite{Mesaros2018_DCASE} recorded in Europe and the CochlScene dataset \cite{jeong2022cochlscene} recorded in Korea. Following the DNS2020 data-generation procedure, 100 hours of noisy-clean speech pairs are synthesized for training.

\noindent \textbullet \hspace{1pt} \textbf{$\mathcal{D}_4$ (VoiceBank):} This domain is constructed using clean speech from VoiceBank \cite{veaux2013voice} and noise recordings from DEMAND \cite{thiemann2013diverse}. Training, validation, and test sets are generated following the protocol of \cite{kuhne2026mambattention} using the DNS2020 data-generation script. 

Base model $\mathcal{M}$ is trained on $\mathcal{D}_0$ and other datasets are learned incrementally in the order $\mathcal{D}_1 \rightarrow \mathcal{D}_2 \rightarrow \mathcal{D}_3 \rightarrow \mathcal{D}_4$.
At each incremental step, the model has access only to the current domain while being evaluated on all previously encountered domains. While the domain order is not expected to substantially affect the domain-aware setting, it may influence the domain-agnostic setting due to the adapter-selection process. We leave the investigation of alternative domain sequences to future work.
We compare the proposed {DIL-GenSE} against three competitive baselines commonly used to assess the generalization and adaptation capabilities of pretrained generative models:
(1) Generalization (GRL): We evaluate the out-of-domain generalization capability of the pretrained base model $\mathcal{M}$ 
without any adaptation to the incremental domains. (2) Fine-tuning (FT): For each incremental domain, we adapt the $\mathcal{M}$ incrementally by fine-tuning only the linear layers, namely the output prediction head and the WavLM projection layer, while keeping all other model parameters frozen. (3) Joint fine-tuning (Joint FT): For each incremental domain, the linear layers of the $\mathcal{M}$ are fine-tuned using data from all domains encountered up to the current learning stage. Although it violates the incremental learning setup, we included it for completeness.  

\subsection{Implementation details and evaluation metrics}
The base model $\mathcal{M}$ is trained on $\mathcal{D}_0$ with a batch size of 256 for approximately 185k optimization steps, using AdamW optimizer with a learning rate of $5\times10^{-5}$ and 1,000 warmup steps.
For domain-incremental adaptation, LoRA with rank $r=8$ is inserted into the GPT-2 attention and feed-forward layers, while bias parameters remain frozen. The LoRA parameters are trained for 10 epochs with a learning rate of $2\times10^{-4}$ and a cosine learning-rate schedule with 200 warm-up steps. For FT, we set the learning rate to $1\times10^{-3}$, keeping other settings the same as LoRA.

Following the standard practice in incremental learning \cite{mcdonnell2023ranpac}, we evaluate the performance of each method after learning the current domain $\mathcal{D}_t$ using the average SpeechBertScore \cite{saeki24_interspeech} (SBS) and average forgetting over all domains encountered so far. The average SBS is defined as:  $\text{SBS}_t = \frac{1}{t}\sum_{j=1}^t \text{SBS}_{t, j}$, where $\text{SBS}_{t, j}$ is the SBS of $j$-th domain after learning the $t$-th domain. Based on this, average forgetting is  defined as: $ \text{FR}_t = \frac{1}{t-1} \sum_{j=1}^{t-1} \left( \max_{\hat{t}\in\{1,\ldots,t-1\}} \text{SBS}_{\hat{t},j} - \text{SBS}_{t,j}\right),$
where $\text{SBS}_{\hat{t},j}$ represents the best (highest) SBS achieved on domain $\mathcal{D}_j$ before learning the current domain. Lower forgetting values indicate better retention of previously acquired knowledge. We additionally report DNSMOS \cite{reddy2022dnsmos} scores, computed using the toolkits and settings provided in \cite{yao2025gense},\cite{li2025sense}. These metrics are also used to compare our DIL-GenSE with the state-of-the-art generative SE model GenSE \cite{yao2025gense}.

\subsection{Results}

\begin{table}[]
 \caption{Average $\text{SBS}_t$ ($\uparrow$) across the current domain $\mathcal{D}_t$ and all previously seen domains $\mathcal{D}_{1,\ldots,t-1}$ under the domain-aware setup.}
 \vspace{-5pt}
     \centering
\resizebox{.8\columnwidth}{!}{%
   \begin{tabular}{l|cccc}
 \toprule
  Method & \makecell{$\mathcal{D}_1$\\DNS} & \makecell{$\mathcal{D}_2$\\EARS} & \makecell{$\mathcal{D}_3$\\LibriTTS-R} & \makecell{$\mathcal{D}_4$\\VoiceBank}  \\
 \midrule
  GRL & 0.89 & 0.82 &0.84 &0.83   \\
\midrule
  FT & 0.85 & 0.78 & 0.74 & 0.73\\
 Joint FT & 0.85 & 0.80 & 0.82 & 0.83\\
 \midrule
  DIL-GenSE & \textbf{0.90} & \textbf{0.87} &\textbf{0.88} &\textbf{0.88}\\
 \bottomrule
 \end{tabular}%
}%
     \label{tab:Aware}
     \vspace{-10pt}
     \end{table}

\begin{figure}[!tbp]
  \centering
  \subfloat[]{\includegraphics[width=0.5\linewidth]{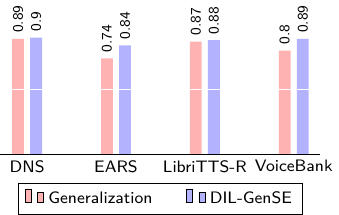}\label{fig:gen_dil_gense}}
  \hfill
  \subfloat[]{\includegraphics[width=0.5\linewidth]{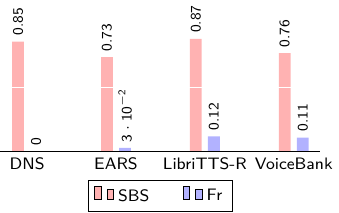}\label{fig:gen_dil_FT}}
  \vspace{-5pt}
  \caption{{Comparison of the proposed DIL-GenSE with zero-shot generalization and fine-tuning. (a) SBS ($\uparrow$) of the generalization and the DIL-GenSE method in the current domain $\mathcal{D}_t$. (b) SBS ($\uparrow$) at the $\mathcal{D}_t$ and average forgetting $\text{FR}_t$ over the previously encountered domains $\mathcal{D}_{1,\ldots,t-1}$ learned for the FT. }}
     \label{fig:compare}
     \vspace{-10pt}
\end{figure}

\textbf{Domain-aware:} The average performance ($\text{SBS}_t$) of the proposed DIL-GenSE is compared with the baselines after learning each domain $\mathcal{D}_t$ in Table~\ref{tab:Aware}. For a more detailed analysis, Fig.~\ref{fig:gen_dil_gense} presents the SBS achieved by each method on individual domains before averaging.
 The pretrained base model $\mathcal{M}$ has previously seen DNS2020 noise recordings during training, which are also used to construct $\mathcal{D}_1$. As a result, $\mathcal{D}_1$ exhibits a smaller distribution shift relative to the pretraining data, leading to better zero-shot generalization performance and a higher SBS compared with the remaining domains. The lower SBS of the GRL on the EARS domain suggests a greater domain mismatch between EARS and the pretraining data compared to the other domains.
 This observation highlights the need for domain-specific adaptation under substantial domain shifts. 
 The introduction of LoRA adapters enables $\mathcal{M}$ to capture domain-specific acoustic characteristics while preserving the domain-shared knowledge encoded in the frozen base model, thereby outperforming the GRL.

We further compare the proposed method with the fine-tuning baseline (FT) in Table~\ref{tab:Aware} and Fig.~\ref{fig:gen_dil_FT}. Fine-tuning the linear layers of $\mathcal{M}$ on the DNS domain overwrites part of the model's pretrained generalization capability, leading to a reduced SBS. 
Interestingly, the SBS achieved by FT on LibriTTS-R is comparable to that of the GRL baseline. This observation suggests that the distribution shift between LibriTTS-R and the previously encountered domains is relatively small. Consequently, adapting the linear layers introduces only a limited bias in the learned feature representations, resulting in performance similar to that of GRL. A similar trend can be observed for DIL-GenSE, whose SBS on LibriTTS-R is also close to that of GRL. This indicates that the contribution of the domain-specific LoRA parameters is less significant when the target domain closely resembles the domains already represented by the base model.

However, once the model is fine-tuned on $\mathcal{D}_3$ (LibriTTS-R), the SBS on the previously learned domain $\mathcal{D}_2$ (EARS) decreases substantially, leading to a higher forgetting score ($\text{FR}_t$), as shown in Fig.~\ref{fig:gen_dil_gense}. This behavior is mainly due to the larger distribution mismatch between the EARS and LibriTTS-R domains, which causes the fine-tuned linear layers to become biased toward the current domain at the expense of previously acquired knowledge. A similar trend is observed after adapting to the VoiceBank domain, where the FT baseline performs worse than both GRL and DIL-GenSE.

Domain-specific LoRA adapters in DIL-GenSE effectively adapt to highly mismatched domains while preserving performance on previously learned domains, thereby outperforming both GRL and FT. In contrast to FT, which is susceptible to forgetting, DIL-GenSE retains domain-shared knowledge in the frozen base model and learns only domain-specific adaptations through lightweight LoRA parameters. As expected, joint FT benefits from access to data from all previously encountered domains and consequently achieves better performance than FT. However, combining data from multiple domains provides only limited gains, as the model tends to become biased toward the dominant domains in the training set. Interestingly, despite having access only to the current domain during adaptation, DIL-GenSE consistently outperforms joint FT, demonstrating its ability to effectively balance domain adaptation and knowledge retention in the domain-incremental learning setting.

\begin{table}[] 
 \caption{Average $\text{SBS}_t$ ($\uparrow$) across the current domain $\mathcal{D}_t$ and all previously seen domains $\mathcal{D}_{1,\ldots,t-1}$ under the domain-agnostic setup.}
     \centering
\vspace{-10pt}
\resizebox{.6\columnwidth}{!}{%
   \begin{tabular}{cccc}
 \toprule
   \makecell{$\mathcal{D}_1$\\DNS} & \makecell{$\mathcal{D}_2$\\EARS} & \makecell{$\mathcal{D}_3$\\LibriTTS-R} & \makecell{$\mathcal{D}_4$\\VoiceBank}  \\
 \midrule
  0.90 & 0.79 &0.80 &0.76   \\

 \bottomrule

 \end{tabular}     
 }
     \label{tab:Agnostic}
     \vspace{-5pt}
     \end{table}
\textbf{Domain-agnostic:}
The performance of DIL-GenSE in the domain-agnostic setup is reported in Table~\ref{tab:Agnostic} and strongly depends on accurately identifying the appropriate LoRA adapter using the uncertainty measure in Eq.~(\ref{uncertainity}). 
Imperfect adapter selection can lead to suboptimal performance compared with the domain-aware setup.
These results suggest that the primary limitation of the domain-agnostic framework lies in the adapter-selection mechanism rather than the LoRA-based adaptation itself. Nevertheless, DIL-GenSE consistently outperforms the FT baseline, indicating that domain-specific LoRA adaptation remains effective even when the domain identity is unavailable during inference. Therefore, developing more reliable adapter-selection strategies has the potential to further improve domain-agnostic performance.

\begin{table}[] 
 \caption{Comparison of proposed DIL-GenSE with GenSE \cite{yao2025gense} on the EARS domain.}
     \centering
\vspace{-10pt}
\resizebox{.8\columnwidth}{!}{%
   \begin{tabular}{l|cccc}
 \toprule
 & \multicolumn{3}{c}{DNSMOS$\uparrow$}\\
 Method & SIG & BAK & OVL & \makecell{SBS$\uparrow$} \\ 
 \midrule 
GenSE\cite{yao2025gense} & 2.60 & 3.22 &2.14 &0.52  \\
 \midrule
DIL-GenSE Agnostic & 3.39 & 3.99 &3.16 &0.74  \\
DIL-GenSE Aware & \textbf{3.56} & \textbf{4.05} &\textbf{3.31} &\textbf{0.84}   \\

 \bottomrule

 \end{tabular}     
 }
     \label{tab:gense}
     \vspace{-10pt}
     \end{table}
We further compare the proposed DIL-GenSE with the state-of-the-art generative speech enhancement model GenSE \cite{yao2025gense} on the EARS domain in Table~\ref{tab:gense}. A  comparison on the remaining domains would be unfair, as GenSE is pretrained using a mixture of clean speech from the DNS, LibriTTS, and VoiceBank datasets, together with noise recordings from the WHAM! and DEMAND datasets.  Therefore, we select the EARS domain for evaluation, as it provides the fairest comparison between the two methods, despite the presence of WHAM! noise in EARS.  The proposed DIL-GenSE in a domain-aware setup outperforms GenSE across all evaluation metrics.
It is worth noting that GenSE relies on two LM decoders for speech enhancement, whereas DIL-GenSE employs a single decoder together with lightweight domain-specific LoRA adapters. Despite its lower architectural complexity, DIL-GenSE consistently outperforms GenSE on the highly mismatched EARS domain in both the domain-aware and domain-agnostic settings.

\section{Conclusion}
We presented DIL-GenSE, a domain-incremental learning framework for generative speech enhancement. DIL-GenSE employs lightweight domain-specific LoRA adapters to enable incremental learning on new domains while mitigating catastrophic forgetting. Each LoRA adapter contains only 1.18M parameters, about 1.1\% of the base model's. Experimental results show that DIL-GenSE consistently outperforms the baseline methods in the domain-aware setting while requiring only a small number of additional parameters. Although DIL-GenSE also improves over fine-tuning in the domain-agnostic setting, its performance remains dependent on accurate adapter selection. Improving domain-agnostic adaptation therefore constitutes an important direction for future research.

\bibliographystyle{IEEEbib_ver1}
\bibliography{strings,reference}

\end{document}